# Seamless Gradient Pixels for Multidimensional Optical Information Interrogation

*Zhenchao Liu*[1], *Bochang Wu*[1], *Houxin Fan*[2], *Tingbiao Guo*[2*], *Hongtao Wang*[3], *Lei Chen*[1], *Cheng-Feng Pan*[1], *Shutao Zhang*[4], *Md Abdur Rahman*[1], *Wen Siang Lew*[5], *Hongyoon Kim*[6,7], *Heechang Yun*[8,9], *Junsuk Rho*[6,7,8,9,10*] and *Joel K.W. Yang*[1*]

1. Engineering Product Development, Singapore University of Technology and Design, Singapore, 487372, Republic of Singapore
2. Centre for Optical and Electromagnetic Research, College of Optical Science and Engineering, Zhejiang University, Hangzhou, 310058, People's Republic of China
3. Beijing Engineering Research Center of Mixed Reality and Advanced Display, School of Optics and Photonics, Beijing Institute of Technology, Beijing 100081, China
4. Quantum Innovation Centre (Q.InC), Agency for Science Technology and Research (A*STAR), 2 Fusionopolis Way, Innovis #08-03, Singapore 138634, Republic of Singapore
5. School of Physical and Mathematical Sciences, Nanyang Technological University, 21 Nanyang Link, 637371, Republic of Singapore
6. Department of Mechanical Engineering, Pohang University of Science and Technology (POSTECH), Pohang, Republic of Korea
7. National Institute of Nanomaterials Technology (NINT), Pohang, 37673, Republic of Korea
8. Department of Chemical Engineering, Pohang University of Science and Technology (POSTECH), Pohang, Republic of Korea
9. Quantum Information Science Program, Graduate School of Convergence Science and Technology, Pohang University of Science and Technology (POSTECH), Pohang, Republic of Korea
10. Department of Electrical Engineering, Pohang University of Science and Technology (POSTECH), Pohang, Republic of Korea

**Abstract:** Information from light interaction with its environment can be extracted from its numerous properties. Yet capturing and interpreting this information within a single and compact platform remains challenging because conventional systems rely on discrete components and detection channels. Here we introduce the concept of "gradient pixels" that provides a universal metasurface platform to interrogate optical field information in the infrared (IR) spectrum. By continuously varying a design parameter across two dimensions, we encode multidimensional optical information into measurable spatial patterns, e.g. both spectral and polarization information are acquired from a single snapshot image. A free-form gradient pixel further enables prescribed wavelength-selective spatial patterns within a single metasurface through tailored gradient profiles. Unlike isolated filter arrays, seamless gradient-pixels multiplex neighbouring meta-atoms, increasing channel spatial density and relaxing $Q$-factor constraints. A 250-μm footprint of these meta-atoms achieves spectral resolution better than 100-pm over a 100-nm spectral span based on a serpentine configuration. Snapshot wavelength-polarization reconstruction yields polarization angular accuracy at ~0.3°. We extended this concept to perform near-field refractive index sensing at video rates using the same explicit parameter-space mapping. By linking far-field detection and near-field sensing, it enables spatially resolved analysis of optical fields and local dielectric environments through a common image-based readout. It offers prospects for multidimensional optical sensing, computational imaging and intelligent information processing.

Light fields are inherently suited to carry information across multiple degrees of freedom, e.g. wavelength, polarization and intensity. Accessing and interpreting this multidimensional information is fundamental to modern optical science and underpins a broad spectrum of applications, including optical metrology and sensing, biomedical imaging, environmental monitoring and optical communications[1-9]. Recent advances in hyperspectral, polarization, and computational imaging have expanded our capability to capture, reconstruct, and interpret multidimensional optical information[10-16]. Spectral acquisition uses dispersion, filtering or interference to encode wavelength information into measurable responses[17,18]. Similarly, polarization imaging has progressed from sequential modulation using rotating polarizers or liquid crystals to the snapshot schemes based on micro-polarizer arrays, interferometry and computational reconstruction[13,19-23]. However, conventional implementations often employ complex wavelength-selective elements or temporal scanning for high-resolution measurements, slowing acquisition and increasing system complexity[24-26]. Snapshot and computational approaches improve acquisition efficiency in drastically more compact systems[1,27-31], although some rely on pre-trained heavy neural networks to recover spectral and polarization information from encoded measurements.

Transitioning from pixelated nanophotonic arrays to continuously varying metasurfaces offers a route to jointly encode multiple optical parameters within a single nanophotonic device. The geometry-dependent resonances of the nanophotonic platforms supporting bound states in the continuum (BIC), photonic crystals and other subwavelength resonators provide control over spectral, polarization and wavefront responses, as well as providing high sensitivity to local dielectric perturbations associated with biological or chemical processes[32-37]. In conventional pixelated implementations, however, each pixel typically comprises an array of identical nanostructures with a predefined optical response. Regarding multidimensional optical information, most optical architectures, however, are still based on discrete sampling and individual sensing functionality, requiring cascaded optical components or separate measurement schemes, leading to increased system complexity and reduced acquisition efficiency. As more dimensions are incorporated, jointly encoding these parameters while maintaining distinguishable responses and tractable reconstruction become challenging, highlighting the need for a compact, integrated architecture for unified multidimensional optical interrogation. Recent advances in continuous gradient metasurface enable spatial readout of optical parameters[38-40], yet a unified architecture that links far-field detection with near-field sensing for spatially multidimensional optical interrogation remains lacking.

Here, we introduce a concept of “gradient pixel” for converting multiple optical parameters of the light field into intensity distributions across symmetry-broken metasurfaces. These metasurfaces then act as microprobes that can be arranged in an array for imaging, analyte analysis and near-field mapping, e.g. of refractive index shifts. As shown in Figure 1a, unlike conventional filter arrays relying on discrete spectral sampling, the seamless spatial mapping between optical parameters ($\lambda$, $\psi$, $n$) and imaging intensity distributions enables direct visualization and decoding of these parameters within a compact imaging architecture. As illustrated in Figures 1b(i-ii), each gradient pixel can jointly retrieve multiple optical parameters, enabling an array of such customized pixels to map their spatial distributions in an image plane or analyze a homogenous beam of light. The same framework further extended to near-field sensing (contact sensing) based on transmissive configuration, where local refractive index variations are transformed into spatially resolvable optical intensity pattern. By unifying far-field light field interrogation (non-contact) and near-field optical sensing (contact) within a single platform (Figure 1b(iii)), our approach provides a universal route toward multidimensional optical information acquisition across different length scales.

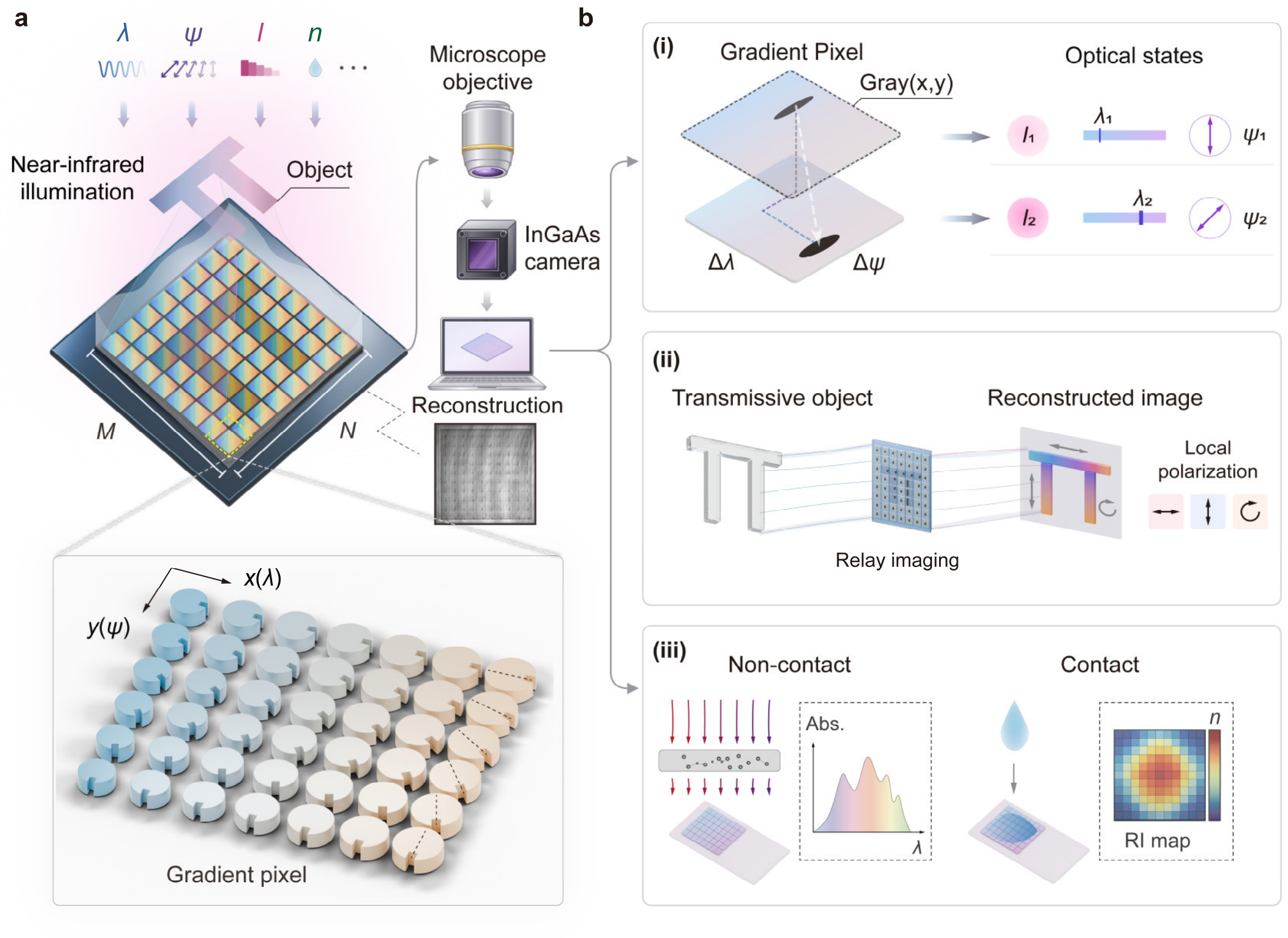

**Figure 1. Concept and principle of the seamless gradient-pixels metasurface platform for multidimensional optical information interrogation.** **(a)** Schematic of the gradient-pixels nanophotonics platform enabling the detection of multidimensional optical information. Here, $\lambda$, $\psi$ and $I$ denote the wavelength, linear polarization angle and relative spectral intensity, respectively, whereas $n$ denotes the refractive index of local dielectric environment. The $\pi$-shaped pattern represents an object imaged onto the gradient-pixel array under near-infrared transmission illumination. Continuously varying meta-atoms, with size and orientation angle gradients along $x$ and $y$, respectively (inset). **(b)** Workflow for multidimensional optical information encoding, decoding and sensing. (i) Representative

image responses generated by different incident optical fields and the corresponding reconstructed intensity, wavelength and polarization angle. (ii) Encoding and reconstruction of spatial optical-field information. The spatial distributions (e.g. a $\pi$-shaped transmissive object) of wavelength, polarization and intensity in the incident field (left) are encoded by the gradient-pixel metasurface (center) and reconstructed from its image response (right). Colors and arrows indicate wavelength and local polarization, respectively. (iii) Multidimensional optical information interrogation functionalities enabled by a single gradient-pixels chip, including the far-field detection of absorption spectroscopy and the near-field imaging sensing of local dielectric environment.

## Design and Characterization

A quasi-bound states in the continuum (BIC) metasurface is designed as the fundamental block for optical field encoding. As illustrated in Figure 2a, the unit cell is the notched symmetry-broken silicon nanopillar on a silica substrate (Supplementary Section 1). The in-plane symmetry is intentionally broken by introducing notches of different sizes (length $a$ and width $b$ in Figure 2a), thereby tailoring the resonance characteristics such as quality factor (Figure 2b), while the pillar radius $r$ and in-plane orientation angle $\theta$ (Figure 2a) can tune the resonance wavelength and polarization features. This notched silicon geometry is selected for its high-index contrast, low near-infrared loss and readily tunable symmetry breaking, enabling high-Q resonance with precise geometric control. We characterized the geometries and Q-factor features of this notched-silicon BIC metasurface (details in Supplementary Section 2). The symmetry perturbation enables radiation leakage from a magnetic dipole-dominated quasi-BIC, as shown by the multipole analysis (details in Supplementary Section 3). The variation of the structural scaling factor $\eta=r/r_0$ induces a shift of the resonance spectra (Figure 2c(i)), indicating a deterministic relationship between structural parameter and spectral response. Here, $r_0$ denotes a reference pillar radius chosen for normalization. Here, we illustrate a variation in the radius, though generally we can introduce a gradual variation in any physical geometry.

A "gradient pixel" is formed by arranging these nanopillars on a periodic lattice with prescribed spatial gradient $\Delta\eta$ in the scaling factor or gradients in other structural parameters, such as the orientation angle. Although BICs are nonlocal resonances relying on the collective coupling of many meta-atoms[34,41], the observation of well-defined resonance fringes in Figure 2c(ii) confirms that this BIC resonant behavior remains supported within continuously varying gradient-pixel. Furthermore, the dip in spatial intensity profile (Figure 2c(iii)) shows a strong correlation with the wavelength-domain transmission. The narrow resonance features further suggest that only a limited number of meta-atoms (along the gradient direction) are required to sustain the resonance (details in Supplementary Section 4). This behavior is likely related to the magnetic dipole-dominated nature of the quasi-BIC mode (Supplementary Section 3), whose resonance characteristics are less sensitive to adjacent meta-atoms.

The spectral-geometry mapping in Figure 2c provides the foundation for constructing the micro-gradient array. Unlike conventional filter arrays relying on discrete spectral sampling, our proposed architecture can form a nearly seamless spectral-spatial sampling owing to the continuous gradient distribution. This seamless spectral sampling, together with the spatial multiplexing provided by neighboring meta-atoms, improves spectral resolution within a compact footprint. Specifically, closely spaced wavelengths generate overlapping resonance stripes and therefore share some of the same meta-atoms. This spatial reuse enables dense spectral sampling without requiring a separate filter region for each wavelength channel. We experimentally validate this continuous encoding through spectral reconstruction using a serpentine 1D micro-gradient pixel (Figure 2d(i-iii)). The serpentine configuration (Figure 2d(ii)) extends the measurement range by increasing the encoding length while maintaining the spatial sampling density. We used the $\ell_1$-regularized least squares (LASSO) with box constraints to reconstruct the spectrum from the images (details in Supplementary Section 5). We also examined the geometric design and spectral reconstruction performance of the single-line array (details in Supplementary Section 6). The reconstructed spectra show excellent agreement with the ground truth, with a wavelength resolution below 100 pm (theoretical estimate: 62.5 pm; Supplementary Section 7) and a measurement bandwidth of at least 100 nm. A quantitative comparison of spectral resolution, measurement range and device footprint with representative counterparts[11,12,27,42-45] is provided in Supplementary Section 8. The achievable spectrum resolution is primarily governed by the synergistic interplay between structural gradients and resonant response (Supplementary Section 7). One can achieve high resolution with an estimated $Q$ factors of a few hundred, relaxing the reliance on extreme-$Q$ resonances and stringent fabrication tolerances.

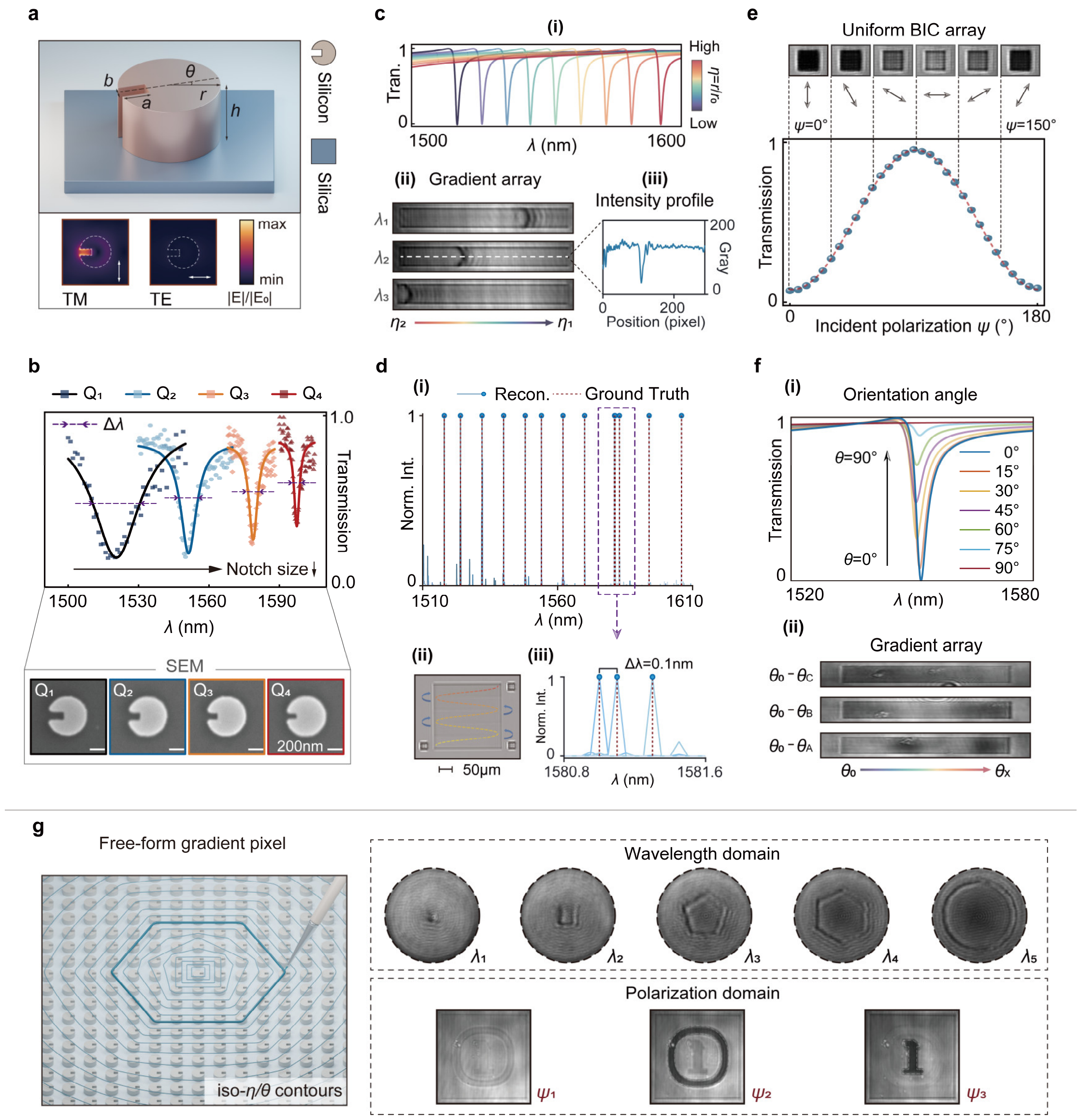

**Figure 2. Design and characterization of the quasi-BIC gradient-pixel metasurfaces for continuous spatial encoding. (a)** Schematic of a unit cell comprising a notched silicon pillar on a silica substrate, with normalized electric-field amplitude, under the TM and TE illumination inset. **(b)** Measured transmission spectra of arrays with unit cell of silicon pillars with different notch geometries ($Q_1$ to $Q_4$) with the pitch of 1 μm and the nanopillar height of 200 nm, illustrating resonance tuning through symmetry breaking. As the notch size decreases, the quasi-BIC resonance redshifts and narrows. **(c)** (i) Simulated transmission spectra for metasurfaces with different structural scaling factors ($\eta=r/r_0$), where $r$ denotes the silicon-pillar radius and $r_0$ is a predefined reference radius used for normalization. (ii) Measured NIR intensity distributions of a gradient pixel with continuous scaling factor encoding ($\eta_1$ to $\eta_2$) under different monochromatic illumination ($\lambda_1$=1507 nm, $\lambda_2$=1537 nm, $\lambda_3$=1559 nm). Here, **t**he designed scaling factor varies from ($\eta_2$=1.0773) on the left to ($\eta_1$=0.9864) on the right. The span of the array is **4**00×50 μm. (iii) Spatial intensity profile along the dashed line in the ($\lambda_2$) image. The intensity minimum marks the position of the resonant dark band. **(d)** (i) Experimental reconstruction of representative incident spectra using the well-designed

micro-gradient pixel. Blue Circles and solid lines denote the reconstructed spectra, while red dotted lines represent the ground truth. Here, the ground truth means the nominal input wavelengths set by the tunable laser, whose output has a very narrow linewidth. (ii) Serpentine gradient-pixel metasurface design for extending the spectral measurement range while maintaining a compact footprint. The dotted colored line traces the serpentine path along which the structural scaling factor varies. Scale bar, 50 μm. (iii) Enlarged view of the region marked in (i), showing the 0.1 nm separation between the two closest peaks. **(e)** Optical micrograph of a 100 μm×100 μm uniform quasi-BIC metasurface array ($Q_1$) under different incident polarization states $\psi$, and the corresponding polarization-dependent transmission measured at the resonant wavelength, showing the modulation of transmission with different incident polarization angles. Double-headed arrows indicate the incident polarization orientations. **(f)** (i) Simulated transmission spectra with different unit-cell orientation angles. (ii) Measured NIR intensity distributions with different orientation angle gradients, which of the spatial intensity contrast follows the trend in the simulated orientation-dependent transmission. $\theta_0$=0°, $\theta_A$=360°, $\theta_B$=180° and $\theta_C$=90°. **(g)** Measured NIR intensity patterns from two free-form gradient pixel. Left, patterns recorded of a single metasurface at five different incident wavelengths ($\lambda_1$-$\lambda_5$), ranging from 1500 nm to 1600 nm, under fixed incident polarization. The wavelength-dependent patterns correspond to selected iso-$\eta$ contours within a continuously varying scale-factor distribution (details in Supplementary Section 10). Right, the image for ($\psi_1$=45°) shows the response at an off-resonant wavelength. The images for ($\psi_2$=0°) and ($\psi_3$=90°) show the response at a second, fixed wavelength. The '0' and '1' patterns are encoded by orthogonal unit-cell orientations.

This resonance also exhibits polarization selectivity, owing to its geometric anisotropy (orientation angle $\theta$ in Figure 2a). The transmission response (Figure 2e) of a uniform BIC array follows a characteristic cosine dependence on the incident polarization state ($\psi$), indicating that polarization information can be mapped into measurable intensity variations (Supplementary Section 9). The simulations in Figure 2f(i) further show that the transmission dip weakens and vanishes as the notched orientation angle increases from 0° to 90°, while the resonance wavelength remains nearly unchanged. Figure 2f(ii) shows the polarization-selective response of the resonance stripes in three orientation-angle-gradient pixels, illuminated by linearly polarized light at resonance wavelength. From top to bottom, the notch orientation spans (0°-90°), (0°-180°) and (0°-360°), with incident polarization angles $\psi$ of 0°, 0° and 90°, respectively. Together with wavelength-selective resonance (Figure 2c), they indicate that the quasi-BIC mode remains robust under geometrical variation.

The shape of resonance stripe is determined by both the parameter of unit-cell and the gradient of array, providing additional degrees of freedom to customized optical-filed responses, including spectrum and polarization, as demonstrated in Figure 2g. The free-form patterns can be encoded through spatial variations in unit-cell scale and orientation. The wavelength-selective patterns are defined by iso-$\eta$ contours within a continuous scale-factor map, whereas the polarization-selective “0” and “1” patterns are encoded using unit cells oriented at $\theta$=0° and 90°, respectively (Supplementary Section 10). The gradient pixel thus enables free-form gradient engineering, serving as a “gradient pen” for sculpting prescribed spatial responses.

**Parameter-Space Mapping Gradient Pixels**

We extend the above encoding strategy for multidimensional light-field interrogation. As illustrated in Figure 3a, an orthogonal dual-gradient pixel, defined by the structural scaling factor ($\eta$) and orientation angle ($\theta$), are integrated into a single metasurface to establish a deterministic dual-parameter mapping from the spectral-polarization parameter space ($\lambda$, $\psi$) to the real-space coordinates ($x$, $y$). This mapping projects independent optical degrees of freedom onto orthogonal spatial dimensions. Simulations (Figure 3b; see Methods) show that each wavelength-polarization combination produces a unique, elongated spindle-shaped intensity pattern within the gradient pixel, allowing distinct optical states to be identified from their spatial responses in real space.

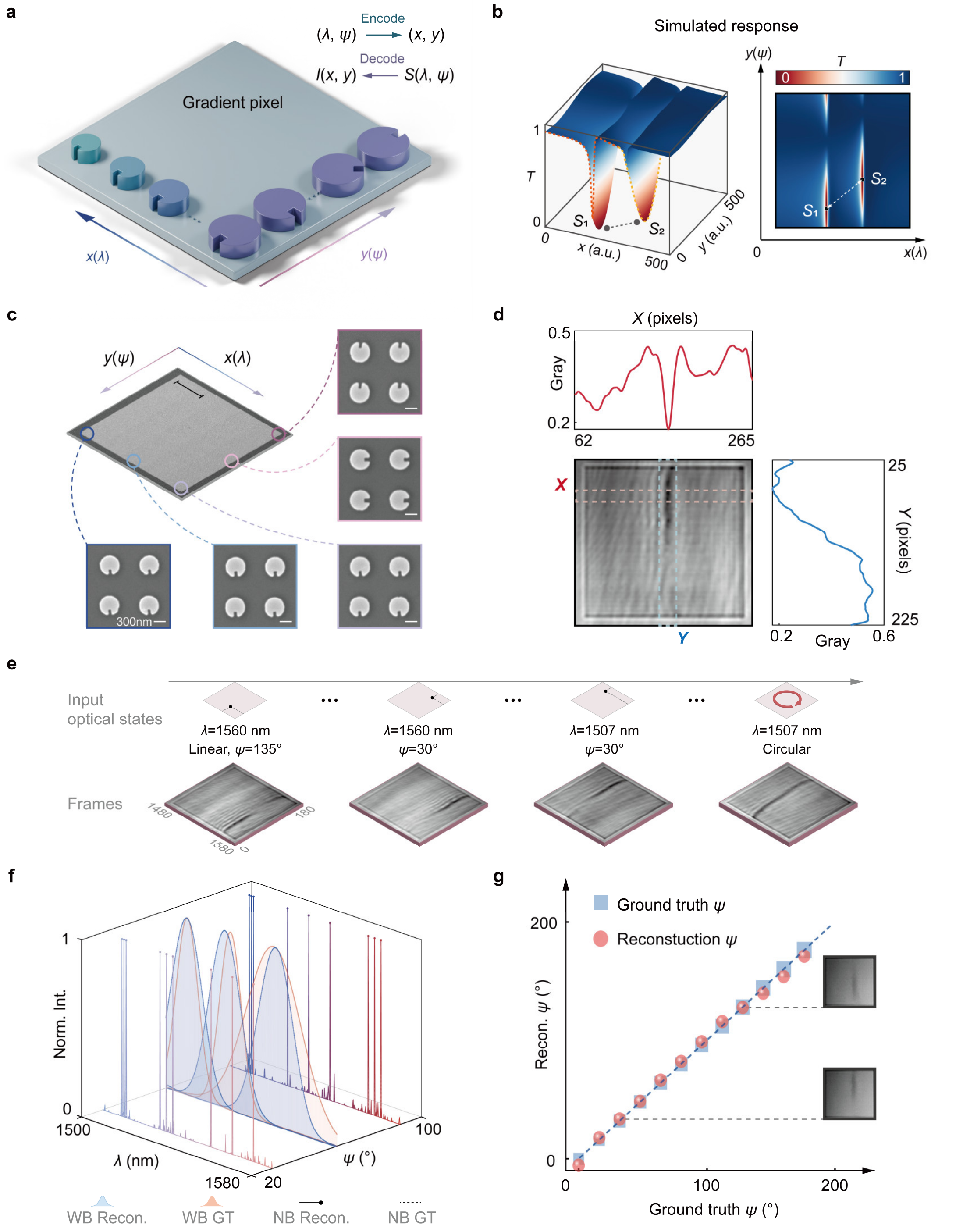

**Figure 3. Mapping principle, device implementation and multidimensional light-field interrogation using a dual-orthogonal gradient pixel. (a)** Schematic of a gradient pixel with spatially varying unit cells on a silica substrate. The silicon-pillar size, controlled by the structural scaling factor $\eta$, varies along $x$, while the in-plane unit-cell orientation angle $\theta$ varies along $y$, establishing a one-to-one mapping between wavelength-polarization space ($\lambda$, $\psi$) and real-space coordinates ($x$, $y$). The measured intensity distribution $I$ ($x$, $y$) is decoded to reconstruct the incident spectral-polarization distribution $S$ ($\lambda$, $\psi$). **(b)** Simulated spatial transmission response $T$ ($x$, $y$) and the corresponding two-dimensional map, showing distinct resonant features ($S_1$ and $S_2$) for two different spectral-polarization states. **(c)** SEM image of the fabricated gradient pixel, highlighting the designed geometric variation along $x(\lambda)$ and $y(\psi)$.

The scale bars in the overview of gradient pixel and in the enlarged unit-cells are 50μm and 300 nm, respectively. **(d)** Measured NIR intensity image and extracted grayscale profiles along the $X$ and $Y$ directions, revealing the wavelength- and polarization-dependent responses respectively. **(e)** Time sequence of input optical states and the corresponding measured NIR intensity images. From left to right, the inputs are 1560 nm with linear polarization at 135°, 1560 nm with linear polarization at 30°, 1507 nm with linear polarization at 30°, and 1507 nm with circular polarization. These changes in wavelength and polarization produce distinguishable spatial intensity patterns. **(f)** Reconstructed broadband (WB) and narrowband (NB) spectra at different linear-polarization angles, compared with ground truth (GT). Broadband inputs are generated using a broadband source and spectral filters. **(g)** Retrieved polarization angles plotted against ground truth under the broadband illumination, exhibiting near-unity correlation and confirming accurate polarization reconstruction.

Figure 3c presents the SEM image of the fabricated gradient pixel with continuous structural variation along both gradient directions. The simulated intensity patterns agree well with the experimental observations in Figure 3d, further validating the proposed parameter-space mapping. Spectrum variations are encoded as spatial intensity distribution along the $x$-direction, whereas variations in polarization state, including both linear and circular polarization states, are independently mapped onto the orthogonal $y$-direction in Figure 3e. We further examined the imaging responses under different incident polarization states (details in Supplementary Section 11). By projecting different optical degrees of freedom onto orthogonal spatial dimensions, spectral and polarization information are naturally decoupled in real space. Consequently, the multidimensional light field can be retrieved from a single intensity image without complex data-driven models. The measured intensity distributions are inversely mapped to the parameter space based on the calibration process, enabling simultaneous reconstruction of the spectral composition (Figure 3f) and polarization state (Figure 3g). The reconstructed results exhibit good agreement with the ground truth in Figures 3f and 3g. Narrowband spectral reconstruction covers an 80 nm wavelength range, with a minimum tested wavelength interval of 1 nm (theoretical estimate: 0.5 nm). Polarization angle reconstruction spans 0-180°, with a minimum error of 0.29°. We used a sinusoidal-based reconstruction algorithm to retrieve the optical parameters from the measured images (details in Supplementary Section 12). The proposed gradient pixel naturally enables snapshot and real-time multidimensional light-field interrogation.

## Multifunctional Optical Information Interrogation Platform

We further develop gradient pixel arrays to detect 2D multidimensional information (Figure 4a). Under transmission illumination, the wavelength and polarization distributions of the optical field modulated by the object, are transformed into intensity patterns. Under transmission illumination, the image of a “SUTD” mask is relayed onto the gradient-pixel array to reconstruct the spatial wavelength distribution (Figure 4b; Supplementary Section 13). The intensity variations within each letter are due to nonuniform illumination. Figure 4c(i) demonstrates the polarization-resolved imaging capability (the “AI” patterns under different polarizations illumination) using the same configuration. For objects with complex spatial structures (Figure 4c(ii); Supplementary Section 14), polarization information at different spatial locations is simultaneously encoded into a single image, enabling quantitative visualization and reconstruction of spatial polarization distributions.

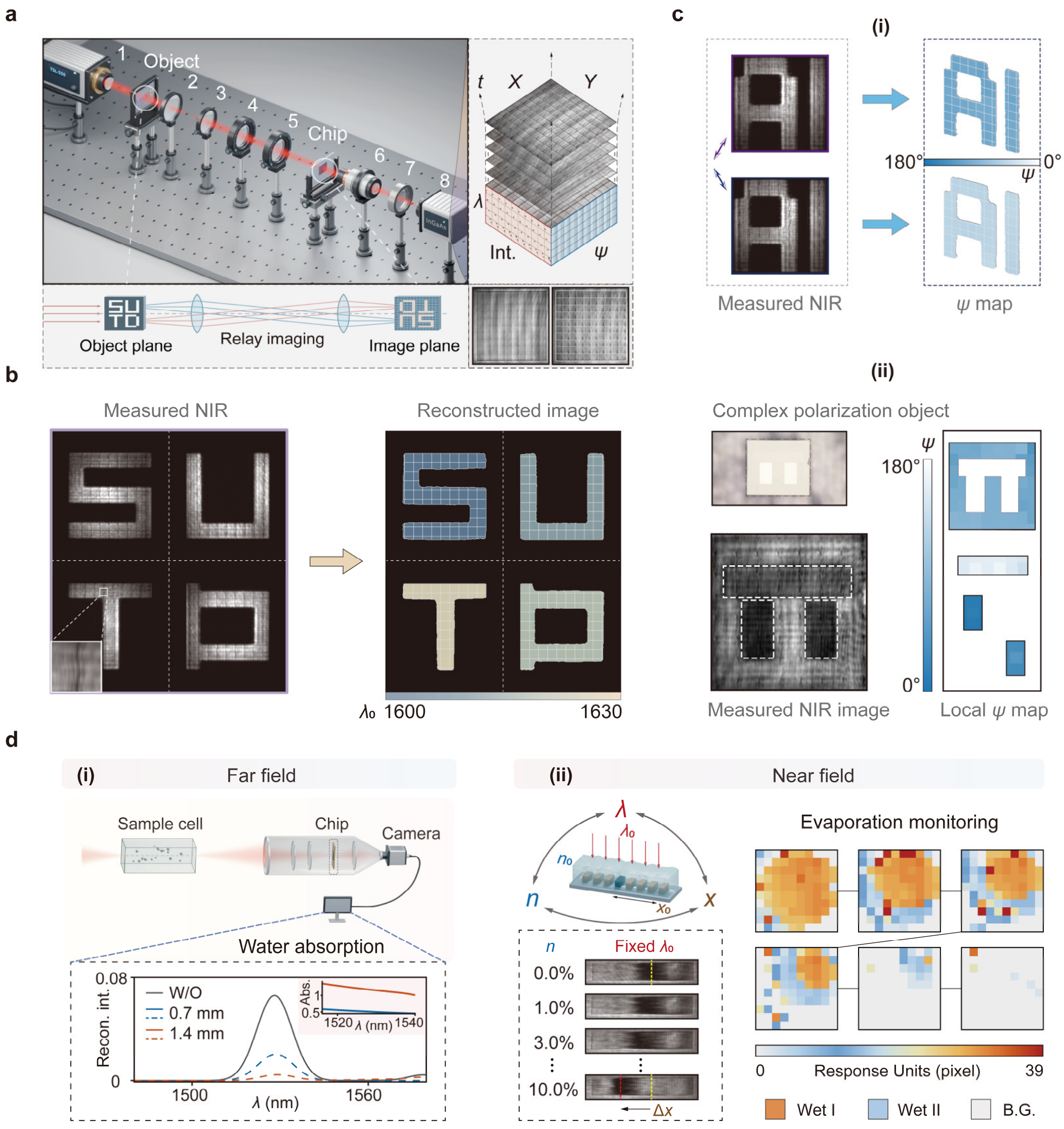

**Figure 4. Unified gradient-pixels platform for multifunctional optical information processing, including spectral, polarization, spatial and environmental dielectric information. (a)** Transmission imaging configuration and data acquisition workflow. A transmissive mask/object at the object plane is relay-imaged onto the gradient-pixel array at the image plane. The mask defines the illuminated regions of the array, and the recorded NIR intensity responses are decoded to reconstruct local wavelength and polarization information. Except for the object and chip, the experimental setup includes (1) Tunable laser; (2, 3) lenses forming a 4f relay system; (4) linear polarizer; (5)

half-wave plate; (6) objective lens; (7) imaging lens; and (8) InGaAs camera. **(b)** Spectrum-resolved imaging of the "SUTD" object/pattern. The captured images are subsequently processed to reconstruct input multidimensional light-field information. The color bar represents the central wavelength of the incident light, varying from 1600 nm to 1630 nm. **(c)** Polarization-resolved imaging. (i) Measured NIR images of the "AI" mask under different incident polarization states and the corresponding reconstructed $\psi$ maps. (ii) Local polarization imaging of a complex polarization object, showing the object photograph, measured NIR image and reconstructed $\psi$ map, including the "$\pi$" pattern. The color bar of $\psi$ varies from 0° to 180°. **(d)** Far- and near-field sensing demonstrations. (i) Far-field water-absorption spectroscopy. The detection of the absorption spectrum of water, including without water (W/O) and for water-path lengths of 0.7 and 1.4 mm. (ii) Near-field dielectric sensing and imaging. At a fixed illumination wavelength $\lambda_0$ (1502 nm), refractive-index changes shift the resonance-stripe position by $\Delta x$. Evaporation response maps, shown chronologically from left to right and top to bottom, reveal a high-response wet state (Wet I, orange label), a lower-response intermediate wet state (Wet II, light blue label) and the dry background (B.G., gray label).

Beyond multidimensional light-field imaging, this platform also supports practical spectroscopic sensing. As shown in Figure 4d(i), the platform operates as a compact spectroscopic sensor capable of measuring absorption spectra. By comparing reference and sample spectra, characteristic absorption features of target media, such as water, can be accurately reconstructed. Due to the sensitivity of the BICs to the surrounding environment, this gradient pixel can be directly extended to near-field sensing and imaging of local dielectric environment in Figure 4d(ii). Variations in the local refractive index perturb the resonance condition of the metasurface and induce measurable shifts in the corresponding spatial response patterns (Supplementary Sections 15 and 16). This relationship (refractive index, resonance wavelength and spatial position ($n$-$\lambda$-$x$)) enables refractive index information to be directly visualized as spatial intensity distributions. We demonstrate both quantitative refractive index sensing and label-free refractive index imaging capable of capturing the spatiotemporal evolution of local dielectric environments, exemplified by the dynamic response of refractive index distributions during water evaporation (Figure 4c(ii); Supplementary Section 17). The response maps show that the high-response region shrinks and nearly disappears as evaporation proceeds, with the array response approaching the background level. It also reveals two coexisting wet states, with the light-blue intermediate state persisting as the high-response region shrinks, which suggests spatially heterogeneous drying across the array. Future designs could use two sensing channels with distinct response sensitivities to disentangle simultaneous variations in incident light and the local dielectric environment. The resulting channel-dependent responses would define a measurement coefficient matrix, enabling the independent reconstruction of the incident-light and refractive index variations.

## Conclusions

In conclusion, we establish a single-chip gradient-pixel platform based on orthogonal dual-gradient quasi-BIC metasurfaces for multidimensional optical measurements. This seamless gradient pixel encodes wavelength and linear polarization into directly measurable spatial intensity distributions, through continuous structural gradients. This architecture spatially multiplexes neighbouring meta-atoms across channels, without requiring a dedicated array for each spectral or polarization channel. With this mapping architecture, single-pixel spectral reconstruction achieves sub-100-pm resolution over a 100-nm range within a 250-μm footprint. Two-dimensional gradient-pixel jointly retrieves wavelength and polarization information from a single NIR microscope image, with a minimum polarization-angle error of 0.29°. Furthermore, the array of gradient pixels extends this approach to spectral and polarization imaging of transmissive objects. The absorption spectroscopy demonstrates the far-field sample characterization. Extending beyond far-field analysis, in the near-field, the same parameter-space mapping converts local dielectric changes into resonance-stripe displacements, enabling label-free, spatially resolved monitoring for liquid evaporation. These results demonstrate how our seamless gradient-pixel be a common measurement strategy to bridge information acquisition across different optical dimensions and sensing regions within one infrared nanophotonic architecture.

## Methods

## Supplementary information

The supplementary information is available online.

## Acknowledgments

## Data availability

The data that supports the findings of this study can be available from the corresponding author upon reasonable request.

## Code availability

The code for reconstructing the spectrum-polarization can be available from the corresponding author upon reasonable request.

## Competing interests

The authors declare no competing interests.